\documentclass{jpp}
\usepackage{graphicx}

\usepackage[utf8]{inputenc}
\usepackage[T1]{fontenc}
\usepackage{amsmath}

\usepackage{graphicx,wrapfig,times}
\usepackage{float,graphicx,caption,sectionbox,fancybox}
\usepackage{amssymb,amsmath}

 \graphicspath{{./EPS/}}

\newcommand{\angstrom}{\text{\normalfont\AA} }
\title{Evolution of an Arched Magnetized Laboratory Plasma 
in a Sheared Magnetic Field}

\author{Kamil D. Sklodowski\aff{1}
  \corresp{\email{kdsklodowski@ucla.edu}},
  Shreekrishna Tripathi\aff{1}
 \and Troy Carter\aff{1}}

\affiliation{\aff{1}Dept. of Physics and Astronomy, University of California, Los Angeles, CA 90095, USA}
\begin{document}

\maketitle

\begin{abstract}
Arched magnetized structures are a common occurrence in space and laboratory plasmas.  Results from a laboratory experiment on spatio-temporal evolution of an arched magnetized plasma ($\beta \approx 10^{-3}$, Lundquist number $\approx 10^{4}$, plasma radius/ion gyroradius $\approx 20$) in a sheared magnetic configuration are presented. The experiment is designed to model conditions relevant to the formation and destabilization of similar structures in the solar atmosphere. The magnitude of a nearly horizontal overlying magnetic field was varied to study its effects on the writhe and twist of the arched plasma. In addition, the direction of guiding magnetic field, along the arch, was varied to investigate its role in formation of either forward- or reverse-S shaped plasma structures. The electrical current in the arched plasma was well below the current required to make it kink unstable. A significant increase in the writhe of the arched plasma was observed with larger magnitudes of overlying magnetic field. Forward-S shaped arched plasma was observed for a guiding magnetic field oriented nearly antiparallel to the initial arched plasma current, while the parallel orientation yielded the reverse-S shaped arched plasma. 

\end{abstract}

\section{Introduction}\label{intro}
\begin{figure}\centering
	\includegraphics[keepaspectratio,width=\textwidth]{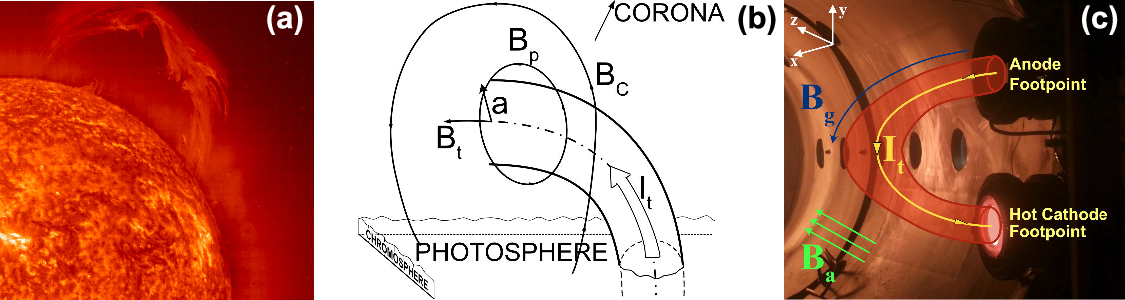}
	\caption{\label{fig01} 
	(a) A solar prominence observed in the extreme ultraviolet (EUV) wavelength of 304 \angstrom  by SOHO spacecraft on 28 March 2000 (Credit$:$ NASA). (b) Schematic of a model flux rope in solar corona with overlying strapping field $B_c$. Minor radius $a$ and flux rope current $I_t$ are indicated. Toroidal and poloidal components of magnetic field are $B_t$ and $B_p$, respectively [adapted from \citep{chen2017}]. (c) Photograph of the experimental setup depicting a current filament $I_t$, guiding magnetic field $B_g$ and ambient (or overlying) magnetic field $B_a$. The experiment simulates background conditions of a solar prominence shown in (b).}
\end{figure}

Arched magnetized plasma structures are ubiquitous in the solar atmosphere. Solar prominences and coronal loops are notable examples of such structures that confine a current carrying plasma by closed magnetic fields. Magnetic loops in the solar corona have characteristically low plasma beta ($\beta \approx 10^{-3} -10^{-2}$) suggesting a nearly force free state of these structures \citep{torok2004, wiegelmann2012, Nakagawa_1973}. A photograph of a solar prominence, a model of its structure, and the laboratory arrangement to model the prominence eruption have been depicted in figure~\ref{fig01}.  
Solar prominences can remain stable for days to few weeks \citep{cambenc, chen_theory_1996, hildner1975, crem2004}. Typical length of these structures is on the order of $10^7$ - $10^8$ m \citep{chen2017}. Some of these structures lose confinement sporadically and erupt due to loss of equilibrium \citep{abbot,rosner1978,burlaga_1988,rust_1994}. The relatively stable pre-eruption phase for these structures lasts for several thousands of Alfv\'{e}n transit times (time taken by the Alfv\'{e}n wave to travel from one to other footpoint of the arched plasma). Solar eruptive events (e.g. coronal mass ejections, solar flares, and jets) are among the most energetic events associated with the eruption of plasma in the solar system. It is estimated that CME eruptions and flares can release $10^{30}-10^{33}$ ergs of energy in the form of kinetic energy of the bulk plasma motion and/or electromagnetic radiation \citep{chen2017}.

The kink instability is a prominent candidate for triggering eruptions on the Sun. Observational signatures of this instability  are usually associated with helical deformations (i.e. writhe and twist) of filaments and prominences \citep{fan2005,torok2005,sturrock2001,sakurai1976}. Improved understanding of processes that lead to eruptions and affect the dynamics of arched magnetized plasmas is of utmost importance for prediction of solar energetic events. A laboratory plasma experiment was designed at UCLA to address this important topic \citep{tripathi_2010}. The experiment facilitates \textit{in situ} measurements on varieties of arched current-carrying magnetized plasma (see figure~\ref{fig01}c). The foot-points of the arched plasma are anchored on cathode and anode, which impose line-tied boundary conditions at both ends of the arch. Plasma parameters in the experiment are appropriately scaled to capture the essential physics of arched plasma on the Sun \citep{tripathi_2010}. Development of writhe and/or twist of a kink-stable arched plasma, in the presence of a strong overlying magnetic field, is the main focus of this paper. Contrary to our intuition, the low-$\beta$ and kink-stable arched plasma in this experiment displays a non-force-free behavior, which will be shown in 3D measurements of electrical-current-density and magnetic-field.

Modeling solar prominences in a laboratory or computer simulation often involves creation of flux ropes - twisted magnetic structures due to significant poloidal magnetic field generated by the toroidal electrical current (see figure~\ref{fig01}(b) for coordinates) \citep{torok2011, fan2005, low2001, chen2011}. A kink-unstable magnetic flux rope with electrical current along externally imposed toroidal magnetic field, $B_t$, is expected to develop writhe due to dominance of self-generated magnetic field, $B_p$. Quantitatively, writhe is a measure of net self-coiling of magnetic field lines and is related to its total torsion (how sharply it is twisting out of the plane of curvature) \citep{torok2014}. In the absence of an overlying magnetic field, writhe of a flux rope is dependent on the magnitude of electrical current and toroidal magnetic field. Therefore, it is also related to magnetic helicity through $H = F^2(T + W)$; where $H$ is the relative helicity, $F$ is the axial magnetic flux, $W$ is the writhe, and $T$ is the number of turns of the field line \citep{berger2006}. The same formula applies to flux ropes with line-tied boundary conditions at both foot-points. Under ideal MHD assumptions (justified for solar prominences), the magnetic helicity is nearly conserved. Therefore, the twist and writhe are closely coupled to each other. The total twist, $\Phi$, can be expressed as \citep{hood1981},
\begin{equation}\label{twist}
	\Phi = \frac{l B_{\phi}(r)}{rB_z(r)} ,
\end{equation}
where $l$ is the length of the flux rope, $r$ is the minor radius, $B_z$ is the axial magnetic field, and $B_{\phi}$ is the azimuthal magnetic field. When the twist exceeds a critical value $\Phi_c$, the system becomes kink unstable and evolves to reduce the curvature of magnetic field lines. This process lowers the net magnetic energy of the system and effectively converts the twist into writhe  \citep{shafranov1957,kruskal1958,freidberg1982,priest1982}. 
For line-tied magnetic arches (aspect ratio = major radius/minor radius $\approx$ 5), the critical value of the twist parameter, $\Phi_c$, was estimated to be $\approx 3.5\pi$ \citep{torok2004}. Although the exact nature of the stored magnetic energy prior to solar eruptions is still a matter of debate, the free magnetic energy stored in a sheared magnetic configuration is identified to be a leading candidate in contemporary computer simulations \citep{fan2005, linker2003,gibson2006}. An association between the sign of magnetic helicity and shape of filaments was suggested by several authors \citep{nakagawa1971}. It has been observed that reverse-S shaped structures dominate the Northern Hemisphere of the Sun, while forward-S shaped structures are more abundant in the Southern Hemisphere \citep{rust1996}. The writhe and twist naturally develop in a sheared magnetic configuration, which produce sigmoidal shaped (forward-S or reverse-S) solar filaments \citep{torok2014}. We note that the total twist in a sheared magnetic configuration may not be accurately estimated using equation \ref{twist} due to underlying assumptions of azimuthal symmetry.

Experimental studies on the arched plasmas have been conducted in the past by a number of groups. The first laboratory experiment simulating the solar arched plasmas demonstrated that an arched magnetic flux rope (AMFR) can be created by driving an electrical current along a guiding magnetic field  \citep{bostick1956}. The next generation of laboratory experiments on arched plasmas were developed in the early 2000s by the Caltech group \citep{bellan1998, hansen2001, hansen2004, tripathi2007}. Their findings include the explanation of sigmoidal shapes and filamentation of the current channels via force-free state equation, and demonstration that the strapping field can inhibit the eruption of solar prominences. The Caltech group also researched on the kink instability, identifying it as a poloidal flux amplification mechanism \citep{hsu_2003}. That work has been followed by AMFR experiments conducted at FlareLab \citep{arnold_2008, soltwisch_2010}. At Princeton, the MRX group investigated arched plasma stability in terms of kink and torus instability parameters \citep{myers2015, myers_2015}, where they identified the guiding magnetic field tension force as the key mechanism to suppress eruption. The MRX group also studied the low-$\beta$ MHD forces in an arched laboratory plasma \citep{myers_2016}. Dynamics of straight magnetic flux ropes, including  magnetic reconnection and kink instability, have been extensively studied in laboratory experiments \citep{gekelman_1992, bergerson_2006, intrator_2009, lawrence_2009, furno_2006}.
In the above-mentioned laboratory research on arched magnetized plasmas, the electrical current rises to several kiloamperes within few Alfv\'{e}n transit times.  As a result, the outward hoop force and strong poloidal twist dominate the arched plasma dynamics during the pre-eruption phase - unlike pre-eruptive solar arched plasmas with less than two poloidal twists  from one foortpoint to the other \citep{Casini_2003, leroy_1983}].  A new approach to laboratory simulations of solar AMFRs was introduced at UCLA to capture the essential features of solar AMFR eruptions \citep{tripathi_2010, tripathi_2013}. Due to much lower electrical current ($<$200 A) and  poloidal twist of magnetic field, the UCLA setup captures essential features of solar arched plasmas during the pre-eruption phase. 
The experiments at UCLA introduce two independent plasma sources, producing the arched magnetized plasma and the background plasma. The evolution of the arched plasma takes place in the presence of a background magnetized plasma, which plays an important role in wave excitation and energy transport. Moreover, the relative magnitude of parameters in the arched and background plasma can be varied, and the magnetic field direction can be reversed. Most importantly, the electrical current in the arched plasma can be kept below kink-instability threshold long enough ($>50 t_A$) to study behavior and evolution of an arched plasma during the pre-eruption phase. The high reproducibility of this experiment and the ability to take measurements in three spatial dimensions allows for reconstruction and visualization of magnetic field, current density and other plasma parameters in three dimensions and in time. 

The development of writhe in a kink-stable arched plasma in a sheared magnetic configuration was studied in this experiment. It is demonstrated that occurrence of the kink-instability is not a necessary requirement for the formation of writhe and twist.  Dependence of the writhe of the arched plasma on the magnitude of the overlying magnetic field has been examined. In addition, the parallel and antiparallel orientations of guiding magnetic field have been correlated with the occurrence of forward-S and reverse-S shaped arched plasmas.

\section{Experimental Setup}\label{experimental}
\begin{figure}\centering
	\includegraphics[keepaspectratio,width=\textwidth]{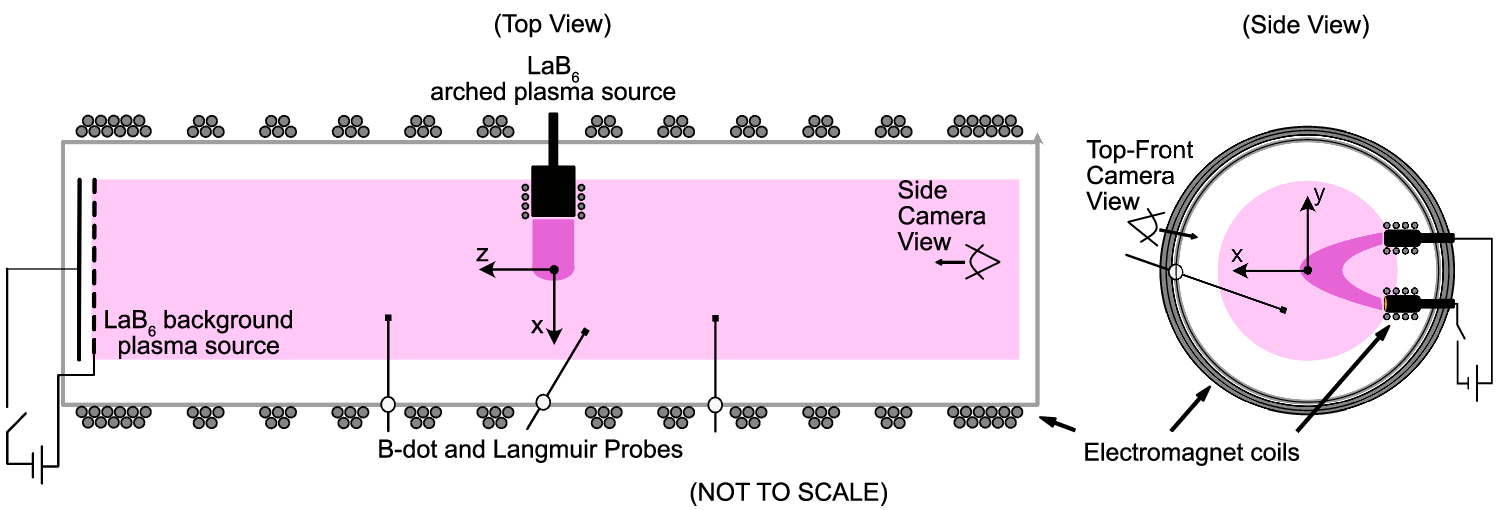}
	\caption{\label{fig02} 
		Schematic of the experimental setup depicting cross-sections of the vacuum chamber from top (left panel) and side (right panel) views. The coordinate system used throughout this paper and its origin are indicated on both panels. The origin is located on the axis of the vacuum chamber, in front of the arched plasma source, as indicated in both panels. The ambient (or background) plasma column is highlighted by a light pink color. The arched plasma is shown in a dark-pink color in both views. These plasma sources are operated in sync using two different discharge sources. The combination of magnetic field generated by larger electromagnets outside the chamber and smaller electromagnets around footpoints of the arched plasma produces a sheared magnetic configuration and provides flexibility in simulating varieties of force-balance scenarios for the arched plasma evolution.
		}
\end{figure}

The experimental setup was designed with a primary goal of modeling the arched magnetized and current-carrying plasmas on the Sun. This is accomplished by driving an electrical current between two electrodes along arched vacuum magnetic field (see figures~\ref{fig01}(c) and \ref{fig02}).
Foot-points of the arched plasma are anchored on electrodes. 

The experiment is performed in a cylindrical vacuum chamber (5.0 m long, 1.0 m diameter). Helium is used as a background neutral gas (pressure: 5 - 9 mtorr). As shown in figure~\ref{fig02}, multiple electromagnets are placed around the vacuum chamber to produce a near uniform and up to 300 Gauss axial magnetic field inside the chamber. The axial magnetic field confines the ambient cylindrical plasma. It also provides an overlying magnetic field for the arched plasma, which impacts the eruption dynamics. The guiding magnetic field (900 G at foot-points) is produced using two smaller electromagnets that surround the cathode and anode (the foot-points of the arched plasma, see the side-view in figure~\ref{fig02}). The ambient plasma is produced by a lanthanum hexaboride (LaB$_6$) hot-cathode source ($\approx$20 cm diammeter).  A discharge is created between this emissive cathode and a molybdenum wire-mesh anode that is located 30 cm away. This plasma source is placed at one end of the vacuum chamber and connected to a discharge pulser (V$_{\text{max}}$: 200 V, I$_{\text{max}}$: 2.5 kA, Repetition rate: 0.5 Hz, pulse-width: 15 ms). The cathode is indirectly heated up to 1700$^{\circ}$C. At this temperature, it becomes efficient in thermionic emission of electrons \citep{cooper2013}. The ambient plasma (0.6 m diameter, 4 m long, plasma density $n_e = 10^{12}$ cm$^{-3}$,  electron temperature $T_e = 4$ eV, pulse-width = 10-15 ms) is produced by acceleration of primary electrons from the cathode during the discharge pulse. The arched plasma (pulse-width = 0.2-0.8 ms) is created using another cathode/anode pair.  The anode is a 15 cm diameter copper disk and the cathode is a 7.6 cm diameter indirectly heated LaB$_6$ disk (temperature $\approx$ 1800$^{\circ}$C).  This anode/cathode pair is mounted on two side ports on the chamber. Typically, the arched plasma has $n_e = 5 \times 10^{13}$ cm$^{-3}$,  $T_e = 13$ eV, Alfv\'{e}n transit time $\tau_A = 2$ $\mu$s, and resistive diffusion time $\tau_R = 500$ $\mu$s.  The arched plasma source uses a floating power supply and it operates in sync with the ambient plasma source. Separation between the foot-points of the arched plasma can be varied in the range of 15 - 41 cm. 
 Both electrodes reside in the $z=0$ symmetry plane, at $x=-28$ cm and $y=\pm$ 13.5 cm (coordinate system depicted in figure~\ref{fig02}).  Typical parameters of the arched plasma and a quiescent solar prominence are presented in table~\ref{table01}. 
\begin{table}
  \begin{center}
\def~{\hphantom{0}}
  \begin{tabular}{lcc}
                    & {Solar Prominence}    & {Laboratory Arched plasma}\\[3pt]
   Plasma $\beta$ 	& 10$^{-2}$ - 10$^{-4}$ & 10$^{-1}$ - 10$^{-3}$     \\
   r/r$_i$			& 10$^9$ - 10$^{10}$    & 10$^2$                    \\
   Lundquist number & 10$^{12}$ - 10$^{14}$ & 10$^3$ - 10$^4$           \\
   Experiment timescale /  $\tau_A$ & 150    & 200                       \\
   Resistive diffusive time / $\tau_A$ & 10$^{10}$ & $>$500               \\
   Aspect ratio 	& 5                     & 3                         \\
  \end{tabular}
  \caption{Comparison of relative plasma parameters of a typical solar prominence \citep{chen2017} and the laboratory arched plasma.}
  \label{table01}
  \end{center}
\end{table}

A computer-controlled 3D probe drive system and a multichannel digitizer are used to acquire high-resolution 3D data on this experiment. Dual-tip Langmuir and three-axis magnetic probes are the main diagnostics tools. These probes are built using high-temperature ceramic coated wires and other components that can withstand up to 750$^{\circ}$C temperature near the cathode foot-point of the arched plasma. This setup allows for a reliable and efficient measurement of plasma parameters ($n$, $T_e$, $B$) with a good spatio-temporal resolution (spatial resolution $\Delta x = 1.5$ cm, temporal resolution $\Delta t = 4\times 10^{-8}$ s for the results reported here). The experiment is highly reproducible, and it operates with a 0.5 Hz repetition rate. This facilitates measurement of the key plasma parameters in 3D. A fast intensified CCD camera (5 ns minimum exposure time, 1280 x 1024 resolution, forced air cooling, and 12 bit digital converter) is used to record the images of the plasma from two different perspectives (side and top-front views as marked in figure~\ref{experimental}).

\begin{figure}
\centering
\includegraphics[keepaspectratio,width=\textwidth]{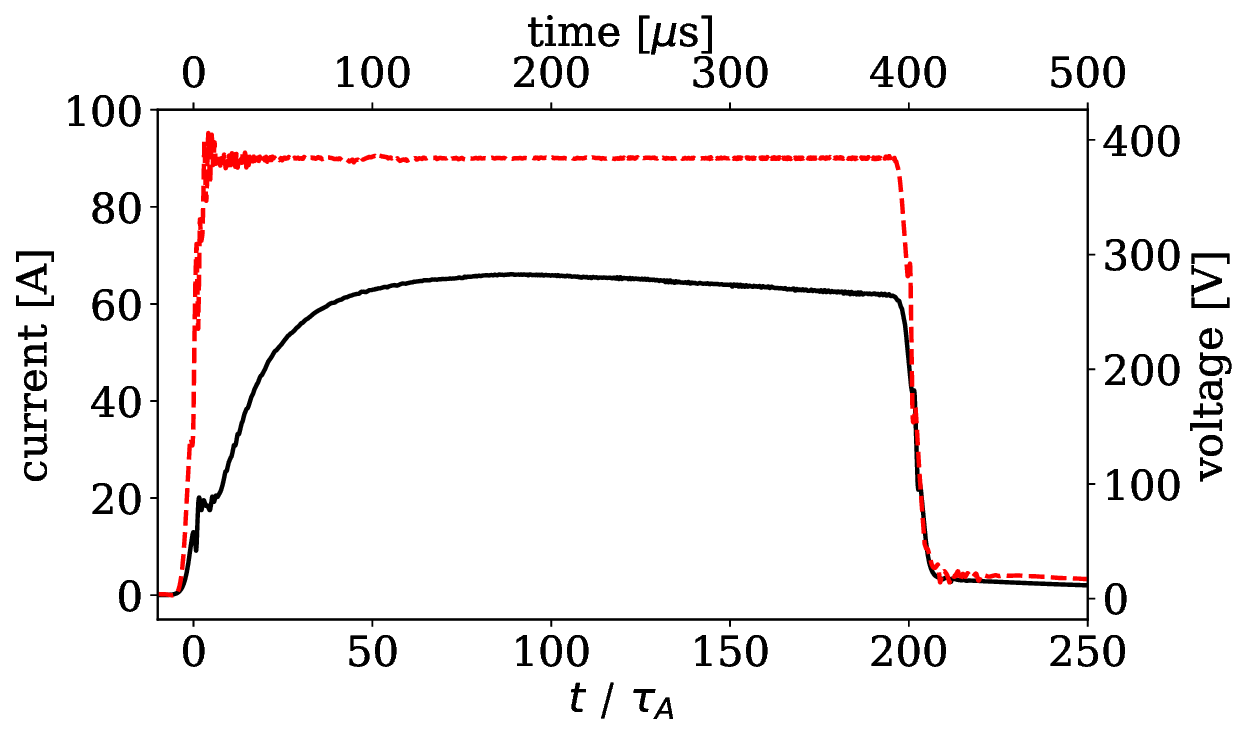}
\caption{\label{Idis} 
		The discharge current (solid black line) and the discharge voltage (dashed red line) time traces of the arched plasma source in the presence of a 15 Gauss ambient magnetic field.
		 Each trace is an average of 16 shots. The guiding magnetic field was oriented parallel to the arched plasma current. Similar trends in the discharge current evolution are observed at other magnetic configurations.  During the first 100 $\mu$s of the discharge, the current gradually builds up and the arched plasma evolves.  A relatively stable phase with persistent appearance of the plasma is observed after $100$ $\mu$s. }
\end{figure}

\section{Results and Discussion}\label{results}
A series of experiments were conducted under different magnetic field configurations to examine the effect of the overlying magnetic field magnitude and the guiding magnetic field direction on the evolution of an arched magnetized plasma. 
The arched plasma current varies in the range of 50-150 A. Typical time traces of the arched plasma current and voltage are presented in figure~\ref{Idis}.
It is evident that the arched plasma current evolves on time scales much faster than the resistive diffusion time ($\tau_R$ $\approx$ 500 $\mu$s).
During earlier stages of evolution ($t < 100$ $\mu$s), the poloidal magnetic-flux in the arched plasma gradually builds up, which leads to dynamic and eruptive behavior. A quasi-steady-state of the arched plasma (nearly persistence appearance with low-frequency global oscillations) is identified at later stages ($t>100$ $\mu$s). 

\begin{figure}\centering
	\includegraphics[keepaspectratio,width=\textwidth]{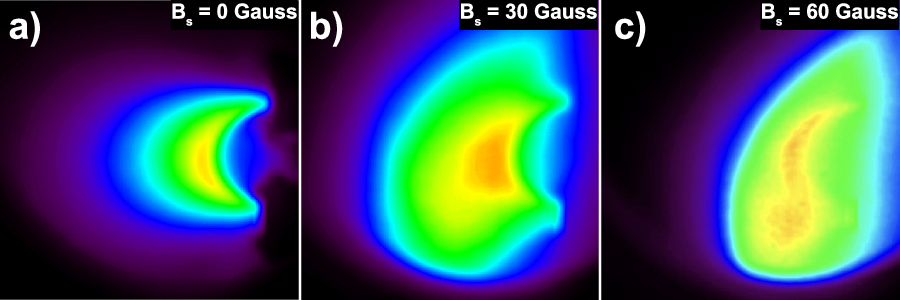}
	\caption{\label{fig04} 
		Unfiltered camera images of the arched plasma taken along z-axis that present the side-view perspective in figure \ref{fig02}. The red and yellow colors signify a higher plasma density, whereas blue and purple color represent the lower density edge region of the arched plasma. Panels (a), (b), and (c) correspond to overlying background magnetic fields of 0, 30, and 60 Gauss, respectively along the positive z-axis (into the page). Each frame was taken at 300 $\mu$s after discharge. These panels represent the final stages of the arched plasma evolution. The earlier stages of the evolution are better captured in 3D magnetic-field data (presented in figure~\ref{fig06}). 
		Deformation of the arched structure and formation of a sigmoid shape are observed at higher ambient magnetic fields in panels (b) and (c).}
\end{figure}

Unfiltered images of the arched plasma are  presented in figure~\ref{fig04}. These images were recorded along the z-axis using the fast camera (located at $z = -3$ m, see figure~\ref{fig02}).
Three panels in this figure were captured at three different overlying magnetic fields  (0, 30, and 60 Gauss) at 300 $\mu$s after the initiation of the arched plasma discharge, which corresponds to the final stage of the arched plasma evolution. Figure~\ref{fig04}a corresponds to a case with no overlying magnetic field, where the plasma evolves to a uniform arch. Significant changes in the morphology of the arched plasma are evident at higher overlying magnetic fields, most notably the appearance of an S-shaped bright region (see figure \ref{fig04}c). The main role of the overlying magnetic field in affecting the dynamics of the arched plasma can be explained by highlighting the importance of Lorentz force associated with the arched plasma current ($I_t$) and the overlying magnetic field ($B_a$). In the solar atmosphere, this force appears due to the interaction of a large-scale strapping field with the electrical current of a prominence or filament \citep{archontis2005, yokoyama1996, yuan2009} (see figure \ref{fig01}b). Relative directions of the overlying (or strapping) magnetic field and the arched plasma (or prominence) current dictate the direction of this Lorentz force. In most cases, this force is in the inward direction on the Sun, and  it assists in inhibiting the prominence eruption. Therefore, the Lorentz force due to the overlying magnetic is arranged to be in the inward direction in the experiment. In addition to altering the balance of magnetic forces, an overlying magnetic field introduces magnetic shear at the leading edge of the arched plasma, which significantly affects the morphology. Formation of the S-shaped structure in figure \ref{fig04}c is a direct result of the development of a strong magnetic shear in the arched plasma.

\begin{figure}\centering
	\includegraphics[keepaspectratio,width=3.5in]{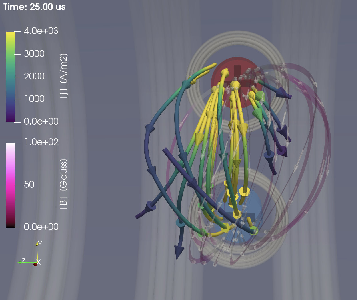}
	\caption{\label{fig05} 
	    Streamlines of the electrical current density of an arched plasma measured at 25 $\mu$s (12.5 $\tau_A$ since the birth of the arched plasma). The overlying magnetic field was turned off in this case. Therefore,  Lorentz force associated with the overlying magnetic field is absent. The solid streamlines (with arrowheads outside the tubes) represent plasma current density, whereas transparent ribbons (with internal arrowheads) represent the total magnetic field (including the vacuum magnetic field). Cathode and anode foot-points are marked by “-” and “+” signs, respectively. Electromagnetic coils are rendered in a transparent light-gray color. The color scales associated with the streamlines of magnetic field and current density are indicated on the left. It is evident that the arched plasma current does not closely follow the vacuum magnetic field lines even in the absence of the overlying magnetic field. 
	}
\end{figure}

\begin{figure}\centering
	\includegraphics[keepaspectratio,width=\textwidth]{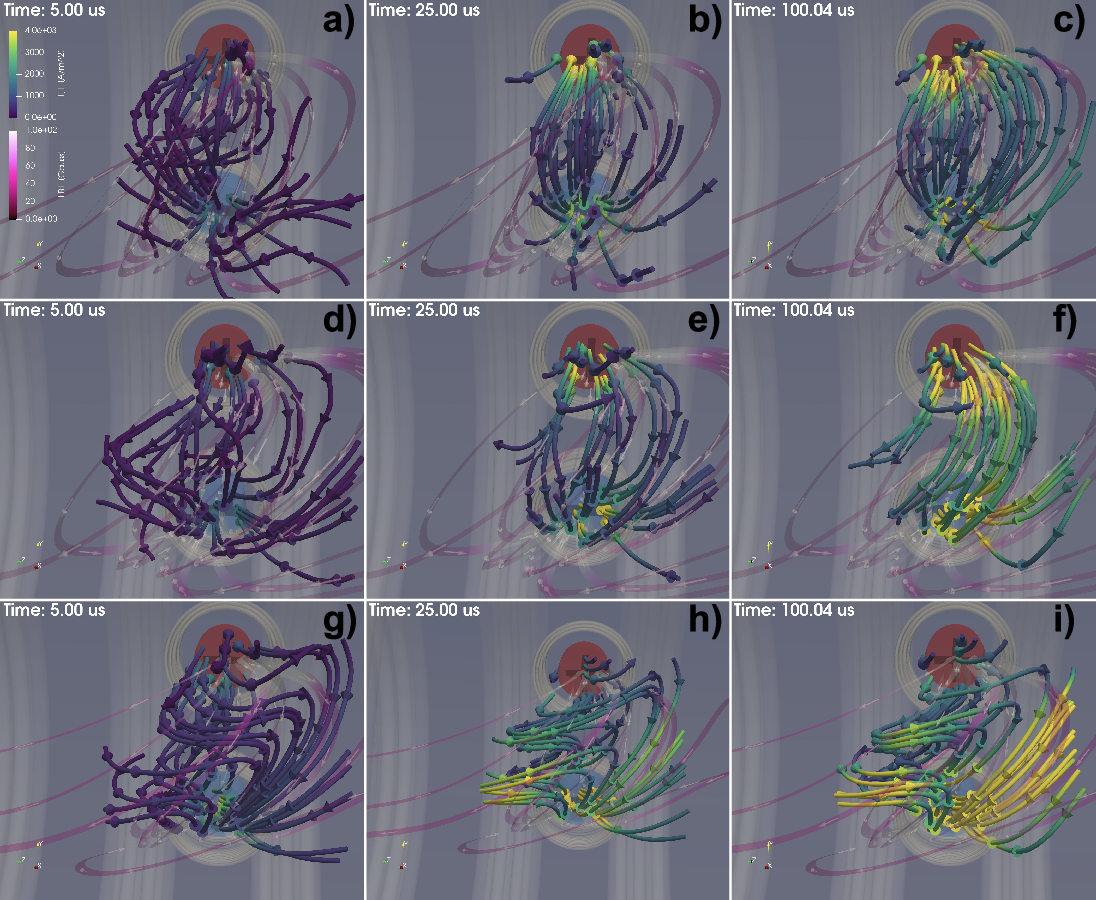}
	\caption{\label{fig06} 
	    The temporal evolution of  the arched plasma is captured by displaying the streamlines of current density and magnetic field at 5 $\mu$s, 25 $\mu$s, and 100 $\mu$s ($\tau_A$ = 2 $\mu$s) for three different ambient magnetic fields: 7.5 G [panels a-c], 15 G [panels d-f], and 30 G [panels g-i]. The guiding magnetic field and the electric current of the arched plasma are nearly parallel to each other in the beginning. The solid streamlines (with arrowheads outside tubes) represent plasma current density, whereas transparent ribbons (with internal arrowheads) represent the total magnetic field (including vacuum magnetic field). Cathode and anode foot-points are marked by “-” and “+” signs, respectively. Electromagnetic coils are rendered in a transparent light-gray color. All panels share the same color scale, displayed on the left of panel (a). The angle between current density lines and the magnetic field lines at the apex increases for stronger ambient magnetic fields. The sigmoid shape of the arched plasma takes a reversed-S orientation. The morphology of the arched plasma at $\approx$100 $\mu$s represents the final stage of the spatio-temporal evolution. 
	    Three movies capturing a high resolution spatio-temporal evolution (-5 to 120 $\mu$s) of the arched plasma for all three magnetic configurations are embedded and uploaded into supplementary materials. 
	}
\end{figure}

\begin{figure}\centering
	\includegraphics[keepaspectratio,width=\textwidth]{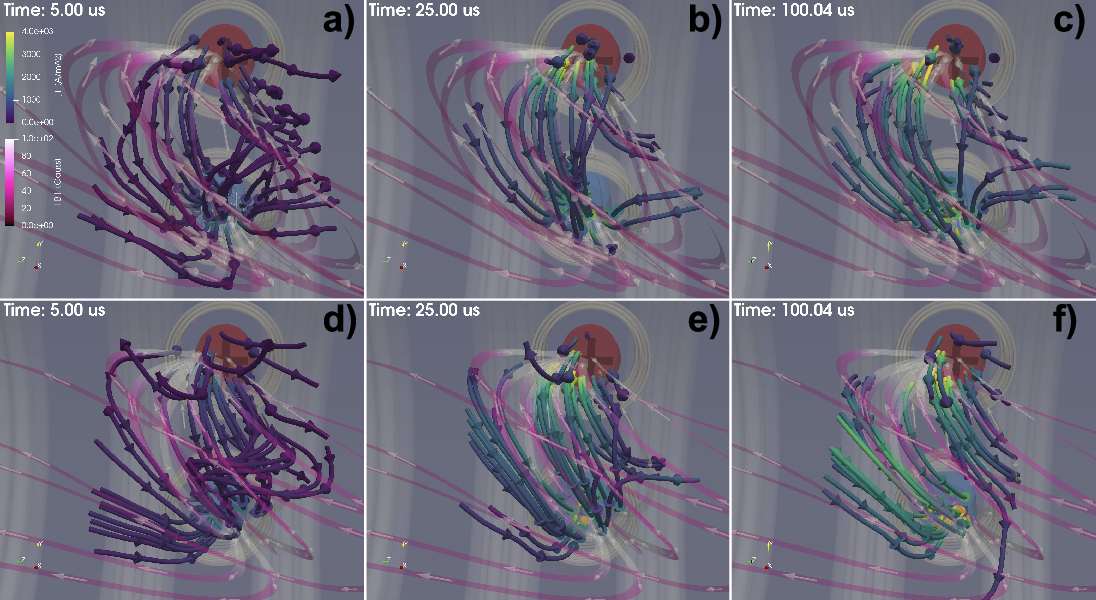}
	\caption{\label{fig07} 
	    The temporal evolution of  the arched plasma is captured by displaying the streamlines of current density and magnetic field at 5 $\mu$s, 25 $\mu$s, and 100 $\mu$s ($\tau_A$ = 2 $\mu$s) for two different ambient magnetic fields: 15 G [panels a-c], and 30 G [panels d-e]. The guiding magnetic field and the electric current of the arched plasma are nearly antiparallel to each other in the beginning. The solid streamlines (with arrowheads outside tubes) represent plasma current density, whereas transparent ribbons (with internal arrowheads) represent the total magnetic field (including vacuum magnetic field). Cathode and anode foot-points are marked by “-” and “+” signs, respectively. Electromagnetic coils are rendered in a transparent light-gray color. All panels share the same color scale, displayed on the left of panel (a). The angle between current density lines and the magnetic field lines at the apex increases for stronger ambient magnetic fields. The sigmoid shape of the arched plasma takes a forward-S orientation. The morphology of the arched plasma at $\approx$100 $\mu$s represents the final stage of the spatio-temporal evolution. 
	    Two movies capturing a high resolution spatio-temporal evolution (-5 to 120 $\mu$s) of the arched plasma for both magnetic configurations are embedded and uploaded into supplementary materials.  
	}
\end{figure}

\begin{figure}\centering
	\includegraphics[width=\textwidth,height=1.5in]{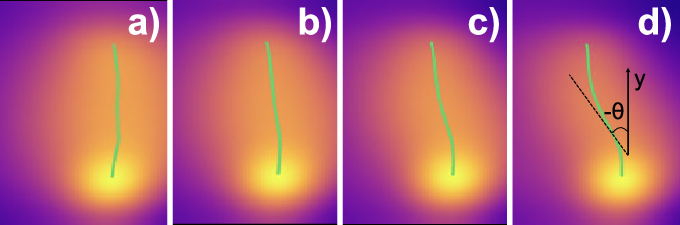}
	\caption{\label{fig08} 
	    Fast camera images of singly ionized helium in the arched plasma (top-front view as indicated in figure~\ref{fig02}, 468 nm narrow pass band filtered) recorded at 125 $\mu$s after the discharge  ($\approx$62.5 $\tau_A$) for (a) 0 G, (b) 7.5 G, (c) 15 G, and (d) 30 G overlying magnetic field configurations. The guiding magnetic field and the electrical current of the arched plasma are nearly antiparallel to each other in the beginning. The symmetry axis of the arched plasma in each frame is highlighted by the solid green line, representing the peak intensity of He$^+$ emission. The magnetic shear of the arched plasma at the apex increases with the strength of the overlying magnetic field. The sheared configuration produces an arched plasma with a sigmoid (forward-S) shape that becomes more pronounced at higher magnitudes of the overlying magnetic field (see panels c and d). This trend is also observed when the guiding magnetic field and the arched plasma current are nearly parallel to each other in the beginning, except for the reverse-S shape of the arched plasma. The angle $\theta$ defined in the text is indicated in panel (d). 
	}
\end{figure}

Images of the arched plasma are useful in identification of global structures and  key stages of the plasma dynamics. However, fine-scale internal structures can be better captured in a high-resolution three-dimensional (3D) measurements of magnetic field. Therefore, analysis of images are complemented by tracking of 3D magnetic-field and current-density structures of the arched plasma for six different magnetic-field configurations. The results for  overlying magnetic-field $B_a$ = 0, 7.5, 15, 30 Gauss with guiding magnetic-field nearly parallel to the initial arched plasma current are presented in figures \ref{fig05} and \ref{fig06}.
Following that, figure \ref{fig07} presents measurements for overlying magnetic field $B_a$ = 15, 30 Gauss with guiding magnetic-field nearly  antiparallel to the initial arched plasma current. The case of $B_a=0$ Gauss is reported to serve as a baseline. The streamlines of plasma current-density and total magnetic-field, in figures \ref{fig05}-\ref{fig07}, are computed by processing volumetric data ($\Delta x = 50$ cm, $\Delta y = 40$ cm, $\Delta z = 40$ cm) from a three-axis magnetic-loop probe. The temporal evolution was recorded with respect to the time when the arched plasma source was turned on at $t = 0$ $\mu$s. Typical Alfv\'{e}n transit and resistive diffusion times are 2 $\mu$s and 500 $\mu$s, respectively. Therefore, changes in the morphology of the arched plasma were minimal after the time indicated in the right-most panels for all magnetic configurations in figures \ref{fig06} and \ref{fig07}. Filtered images of the arched plasma   
were recorded to capture the dynamics of singly ionized helium (468 nm narrow band-pass filter). These images were collected from the top-front view (the camera positioned at x = 65 cm, y = 20 cm, z = 0 cm as indicated in figure \ref{fig02}) for $B_a$ = 0, 7.5, 15, and 30 Gauss. Each case was investigated with the guiding field oriented nearly parallel and then nearly antiparallel to the initial arched plasma current. These images were processed to identify the symmetry axis of the arched plasma. The maximum He-I emission intensity along the horizontal direction was assumed to occur on the symmetry axis. The identification of the symmetry axis is helpful in tracking the morphological evolution of the arched plasma. A selection of processed images is presented in figure \ref{fig08} for the antiparallel orientation of the guiding magnetic field with respect to the arched plasma current.

Relative importance of three major magnetic forces on an arched plasma should be analyzed before looking into its evolution under different scenarios of magnetic field configuration. It is reasonable to assume a half-torus shape of the arched plasma, which is subjected to guiding and overlying magnetic fields (illustrated in figure \ref{fig01}b). Three major magnetic forces on the arched plasma are the tension force, hoop force, and strapping force \citep{chen_theory_1996}. The tension force $F_t$ is a restorative force which tends to decrease the major radius of the arched plasma. It can be expressed as,
\begin{equation}\label{ften}
\mathbf{F_t} = \frac{\hat{R}}{R} \int \frac{1}{\mu_0} \left(B_{gv}^2 - B_{g}^2\right) dS,
\end{equation}
where $\hat{R}$ is a unit vector pointing along the major radius $R$, $\mu_0$ is the permeability of vacuum, $B_{gv}$ is the vacuum guiding magnetic field, and $B_g$ is the total guiding magnetic field (includes contribution from plasma). The hoop force $F_h$ is directed along the major radius, and it assists in the outward expansion of the arch. The hoop force can be expressed as,
\begin{equation}\label{fhoop}
\mathbf{F_h} = \frac{\mu_0}{2\pi R} I^2 \left[ \ln\left(\frac{8R}{a}\right) -1 + \frac{l_i}{2} \right] \hat{R} ,
\end{equation}
where $I$ is the arched plasma current, $a$ is the minor radius, and $l_i$ is the plasma self inductance. 
The third major force on an arched plasma is the strapping force $F_s$. This force  is essentially a Lorentz force between the arched plasma current and the overlying magnetic field. Depending on the direction of the overlying magnetic field, the strapping force can be either inwards (for $B_a$ along positive z-axis) or outwards (for $B_a$ along negative z-axis). The strapping force (per unit length of the arched plasma ) can be estimated by,
\begin{equation}\label{fstrap}
\mathbf{F_s} = -2IB_s\hat{R},
\end{equation}
where $B_s$ is the magnitude of the strapping field. Relative magnitudes of three forces ($\mathbf{F_h}$, $\mathbf{F_t}$, and $\mathbf{F_t}$) govern the dynamics of an arched magnetized plasma in the laboratory and on the Sun. For a typical solar prominence, relative magnitudes of hoop, tension, and strapping forces are 1.0, 0.3, and 0.7, respectively \citep{chen_theory_1996}. The magnetic field configurations were chosen strategically for this experiment to keep it relevant to the solar case. Therefore, it was important to keep the arched plasma current well below the current-threshold for the kink-instability. 

In the case of $B_a$ = 0 Gauss, an outward expansion of the arch was dominated by the hoop force in the absence of an overlying magnetic field and associated Lorentz force (results presented in figure~\ref{fig05}). The twist of the arched plasma for this configuration was observed to be minimal, as expected ($\Phi_{max} \approx 0.05\pi$). The arched plasma was therefore kink-stable ($\Phi < \Phi_c$). Interestingly, the arched plasma current does not closely follow the magnetic field. In the absence of an overlying magnetic field, this magnetic configuration serves as a baseline.
In the following three magnetic configurations, the inward Lorentz force (strapping force) was applied using an overlying magnetic field ($B_a$ = 7.5, 15, 30 Gauss, results presented in figure~\ref{fig06}), while the guiding magnetic field was oriented nearly parallel to the arched plasma current. The overlying magnetic field naturally introduces magnetic shear in the arched plasma. The initial magnitude of the magnetic shear can be enhanced by application of a stronger overlying magnetic field, and its sign can be reversed by reversing the direction of the guiding magnetic field. The magnitude of magnetic shear has a direct impact on the arched plasma evolution. Notably, the sigmoid shape (reverse-S) is prominent at stronger overlying magnetic fields. An increase in the magnitude of $B_a$ enhances the strapping force and reduces the  major radius of the arched plasma. 
For three different overlying magnetic fields, $B_a$ = 7.5, 15, and 30 Gauss, the magnetic field twist  $\Phi_{max} \approx 0.22\pi, 0.35\pi,$ and $ 0.73\pi$, are estimated,  respectively. This confirms that the arched plasma is not kink-unstable. Results in figure~\ref{fig06} suggest that even a kink-stable arched plasma produces a complex magnetic topology in the presence of strong overlying magnetic field. 

Finally, two magnetic configurations associated with antiparallel arched plasma current and guiding magnetic field were explored at $B_a$ = 15, 30 Gauss. Streamlines of magnetic field and electrical current density for both cases are depicted in figure~\ref{fig07}. As expected, the sign of the magnetic shear is reversed in these cases when compared to the parallel guiding magnetic field cases presented in figure~\ref{fig06}. The evolution of the magnetic topology of the arched plasma also differs. The sigmoid shape is again more prominent at stronger magnitudes of overlying magnetic field. However, the arched plasma takes a forward-S shape in this case (as opposed to the reverse-S shape in figure~\ref{fig06}). For both cases, the arched plasma current is well below the current-threshold for the kink-instability. The magnetic field twist is estimated using 3D magnetic-field data. The maximum twist of the arched plasma is  $ 0.22\pi$ and $0.54\pi$ for $B_a$ = 15 and 30 Gauss, respectively. The trend of increase in $\Phi$ at stronger ambient magnetic fields is also observed in images (see figure~\ref{fig08}). 
As discussed in the introduction section, the forward-S and reverse-S plasma structures are usually observed in different hemisphere on the Sun. A close association between sigmoid formation and solar eruption has been established in remote observations \citep{Canfield_2000}. The underlying cause for the formation of forward- and reverse-S shaped sigmoid on the Sun is still unresolved \citep{DeVore_2000}. Our experiments cannot rule out correlation between kink-instability and shape of sigmoid. However, our results confirm that kink instability is not a necessary requirement for the formation of sigmoid on the Sun. In addition, the relative directions of the overlying and guiding magnetic fields play important roles in controlling the sign of sigmoid. 

\begin{figure}\centering
	\includegraphics[keepaspectratio,width=\textwidth]{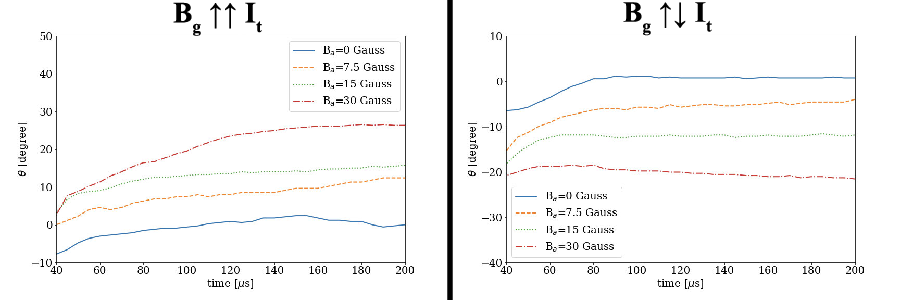}
	\caption{\label{fig09} 
	    Temporal evolution of the shear angle $\theta$ (as indicated in figure~\ref{fig08}d) at four different overlying magnetic fields (0, 7.5, 15, and 30 Gauss) are displayed in these panels. On the left panel (a) the guiding magnetic field is oriented nearly parallel to the initial arched plasma current. The right panel (b) corresponds to the guiding magnetic field, nearly antiparallel to the initial arched plasma current. These panels display results during the later stages of the temporal evolution, mainly because the angle $\theta$ could not be measured reliably in the beginning due to the extremely dynamic nature of the arched plasma. There is a noticeable trend of increase in $|\theta|$ with overlying magnetic field for both guiding field configurations. The negative angle $\theta$ in the antiparallel guiding magnetic field configuration is associated with a forward-S shape, while the positive angle $\theta$ (parallel case) corresponds to the reverse-S shape of the arched plasma.
	}
\end{figure}

Filtered images of singly ionized helium of the arched plasma were captured from the  top-front-view (marked in the right panel of figure~\ref{experimental}) to complement magnetic-field measurements. Cases of $B_a$ = 0, 7.5, 15, and 30 Gauss for both orientations of guiding magnetic field were investigated. Analysis of these images agrees well with estimates of twist from 3D magnetic-field data (depicted in figures~\ref{fig05}-\ref{fig07}). The sigmoid shape is observed to be more pronounced at larger magnitudes of overlying magnetic field, and its sign changes with reversal of the guiding magnetic field. For guiding magnetic-field oriented nearly parallel to the initial arched plasma current, a reverse-S shaped arched plasma is produced. Reversal of the guiding magnetic-field (antiparallel to the arched plasma current) forms a forward-S shaped arched plasma. Selected images for all four cases of ambient magnetic field with guiding magnetic field in the antiparallel orientation are presented in figure~\ref{fig08}. These panels represent final stages of the arched plasma evolution ($\tau_A  << t$ = 125 $\mu$s $< \tau_R$). The solid-green line in these panels highlight the symmetry axis of the arched plasma, where peak emission of He$^+$ occurs in the horizontal direction. The overlying magnetic field significantly impacts the morphology of the arched plasma. The sigmoid shape of the arched plasma is visible at stronger overlying magnetic-field, as evident in figure~\ref{fig08}d.
The temporal evolution of the sigmoid and associated twist are analyzed by acquiring multiple-frames of He$^+$ plasma during 40--200 $\mu$s. 
The temporal resolution of these frames was 5 $\mu$s. 
At each time-step the symmetry axis was computed and marked (as in panels of figure~\ref{fig08}).  The angle $\theta$ quantifies the angle between the vertical axis (connects both foot-points) and the symmetry axis at the leading edge of the arched plasma (marked in figure~\ref{fig08}d). The forward-S shape of the arched plasma is characterized by a negative $\theta$, while the reverse-S shape has a positive $\theta$. The time evolution of $\theta$ for parallel and antiparallel guiding magnetic-field (with respect to the arched plasma current) is presented in figure~\ref{fig09}. The early stages of the arched plasma evolution (0-40 $\mu$s) are excluded from this analysis due to difficulty in accurately identifying the symmetry axis. This was mainly due to lower intensity of He$^+$ emission in the beginning. 
These results are consistent with 3D magnetic-field measurements on  the observation of strong magnetic-shear and sigmoid formation at stronger overlying magnetic fields. Moreover, the reversal of the sign of sigmoid with guiding magnetic field is also confirmed. 

These results indicate that the arched plasma is not in a force-free state (characterized by parallel magnetic field and electrical current). 
There are numerous examples of force-free state of a low $\beta$ plasma - especially in the solar corona. The concept of force-free magnetic fields has been widely used to model the coronal magnetic field structures \citep{Nakagawa_1973}. Although this simplified description is often useful in explaining the large-scale topological evolution of the magnetic fields, it is not always supported by observations - especially during eruptions and in the active region on the Sun \citep{chen2017, McClymont_1997}. The observation of a non-force-free state in this laboratory experiment and active region on the Sun should not be unexpected, since the basic assumptions for the existence of a force-free state (ignorable pressure gradients, plasma flows, and Hall-term) are not supported by observations.
Another example of non-parallel current and magnetic fields are pressure-gradient driven diamagnetic currents in a low-$\beta$ magnetized plasma. In our future research, we will attempt to identify the exact cause of the existence of a non-force-free state in our experiment. In this paper, the main focus is to highlight the development of sigmoid shapes in a kink-stable arched plasma. Our initial analysis suggests eruptive behavior of the arched plasma and formation of large-scale flux-rope structures, when sufficiently large magnetic-shear is developed in the arched plasma ($t = 75-200$ $\mu$s). These large-scale flux-ropes are not captured in figures 6-8 and will be reported elsewhere.

\section{Summary}\label{summary}
The effect of a nearly horizontal overlying magnetic field on the evolution of an arched magnetized plasma has been studied in this laboratory plasma experiment. The experiment was designed to capture the dynamics of arched plasma eruptions on the Sun. The electrical current in the arched magnetized plasma was kept below 200 A to ensure that relative magnitudes of hoop, tension, and strapping forces for solar filaments and prominences are comparable with the experiment. The lower magnitude of electrical current ensures that the arched plasma does not form multiple poloidal magnetic-field twists from one foot-point to the other during the pre-eruption phase (as observed on the Sun).
Experimental results confirm that sigmoid plasma structures are naturally produced in a sheared magnetic configuration – even in the absence of the kink instability. The shear angle critically depends on the magnitude of the overlying magnetic field and the direction of the guiding magnetic field. This suggests that the apparent writhe of a current-carrying arched plasma has a strong dependence on the structure of the overlying magnetic field, not just on the magnitude of the electrical current.  Evolution of the arched plasma in the presence of an overlying magnetic field with a variable decay index will be studied in the near future. The role of magnetic reconnection, Alfv\'{e}n waves, and global oscillations will also be explored using more comprehensive 3D data of plasma parameters.
\\
\\

\textbf{Supplementary data} \\
Supplementary material and movies are available at https://doi.org/10.1017/jfm.2019.  \\

\textbf{Acknowledgments}\\
Authors would like to thank W. Gekelman, and S. Vincena for useful discussions and P. Pribyl, Z. Lucky, M. Drandell, T. Ly, and A. S. Kohli for expert technical assistance. \\

\textbf{Funding}\\
This research was primarily supported by funds from National Science Foundation (Award Number: PHY-1619551). Partial support to S. Tripathi from NASA (HERMES DRIVE Science Center, Award Number: 80NSSC20K0604) is also acknowledged. The experiment is conducted at the Basic Plasma Science Facility (BaPSF) at UCLA, which is supported by US DOE under Contract No. DE-FC02-07ER54918 and the NSF under Award No. PHY1561912.\\

\textbf{Declaration of Interests}\\
The authors report no conflict of interest.\\

\textbf{Author ORCID}\\
K. D. Sklodowski, https://orcid.org/0000-0002-3782-9440; S. K. P. Tripathi, https://orcid.org/0000-0002-6500-2272; T. Carter, https://orcid.org/0000-0002-5741-0495 \\

\bibliographystyle{jpp}

\bibliography{paper}

\end{document}